\documentclass[aps,prf,longbibliography]{revtex4-2}
\usepackage{blindtext}
\usepackage{color}

\usepackage{graphicx}
\usepackage{subfigure}
\usepackage{float}
\usepackage{dcolumn}
\usepackage{bm}
\usepackage{soul} 
\input epsf
\usepackage{sidecap}

\sidecaptionvpos{figure}{c}

\def\u{{\rm \bf u}}

\newcommand\beq{\begin{equation}}
\newcommand\eeq{\end{equation}}

\newcommand{\ks}{\textcolor{black}} 
\usepackage{color} 

\begin{document}
\title{Two-layer channel flow involving a fluid with time-dependent viscosity}
\author{Kirti Chandra Sahu}
\email{ksahu@che.iith.ac.in}
\affiliation{
Department of Chemical Engineering, Indian Institute of Technology Hyderabad, Sangareddy 502 285, Telangana, India}

\begin{abstract}
A pressure-driven two-layer channel flow of a Newtonian fluid with constant viscosity (top layer) and a fluid with a time-dependent viscosity (bottom layer) is numerically investigated. The bottom layer goes through an ageing process in which its viscosity increases due to the formation of internal structure, which is represented by a Coussot-type relationship. The resultant flow dynamics is the consequence of the competition between structuration and de-structuration, as characterised by the dimensionless timescale for structuration $(\tau)$ and the dimensionless material property $(\beta)$ of the bottom fluid. The development of Kelvin–Helmholtz type instabilities (roll-up structures) observed in the Newtonian constant viscosity case was found to be suppressed as the viscosity of the bottom layer increased over time. It is found that, for the set of parameters considered in the present study, the bottom layer almost behaves like a Newtonian fluid with constant viscosity for $\tau>10$ and $\beta>1$. It is also shown that decreasing the value of the Froude number stabilises the interfacial instabilities. The wavelength of the interfacial wave increases as the capillary number increases.
\end{abstract}


\maketitle

\section{Introduction}
\label{sec:intro}

Fluids with time-dependent viscosity are encountered in many processes in hydraulic engineering (e.g. dredging, hyperconcentrated flow, erosion resistance \cite{engelund1984instability,kelly1979erosion}), mining engineering (drilling muds; \cite{nguyen1985thixotropic}), civil engineering (cement and grouts; \cite{lapasin1983flow}), coating technology \cite{weinstein2004coating}, and chemical engineering \cite{ramirez2006modeling,boek2008deposition}. These fluids are classified into two types based on whether their viscosities increase (rheopectic) or decrease (thixotropic) with time. The literature associated with time-dependent fluids and their applications is extensively reviewed in Ref. \cite{mewis2009thixotropy,rg2014}. 

A few researchers have also modelled clay and mud in riverbeds as time-dependent fluids \cite{toorman1997modelling,mitchell1974flowsliding}. The mud accumulated in the riverbed acts as a distinct phase, with clean water flowing over it. The muddy phase ages, which may be caused by a `jamming' transition \cite{coussot2005continuous} of the particulate phase confined within the deposit, resulting in structure-building in the muddy phase (referred to here as `structuration'). For this layer to flow, this structure must be broken (referred to here as `de-structuration'). Thus, it is possible to model the transition from a liquid-like to a solid-like behaviour using a time-dependent rheological model \cite{huynh2005aging,roussel2004thixotropy}. In the present work, a characteristic problem of this kind, i.e. a pressure-driven two-layer channel flow of a Newtonian fluid with constant viscosity and fluid with a time-dependent viscosity \ks{(using a Coussot-type model)} is studied. The development and suppression of interfacial instability have been examined for a wide range of dimensionless parameters associated with this problem. Of course, the continuous deposition of the particulate phase has been neglected in the present study. \ks{Moreover, the applicability of a Coussot-type model to represent the mud dynamics must be experimentally validated.}

Earlier, several researchers have examined the instability in two-layer channel flow involving Newtonian and non-Newtonian fluids because of their importance to practical applications, such as crude oil transportation in pipelines and liquid mixing using centreline injectors, upstream of static mixers, the removal of highly viscous or elastoviscoplastic material adhering to pipes by using fast-flowing water streams, to name a few \cite{joseph97a,rg2014}. For instance, by considering Newtonian fluids, several authors investigated the effect of the ratios of viscosity, density, and thickness of the fluid layers on the instability developed at the interface by conducting linear stability analyses \cite{yih67a,yiantsios88a,hooper83a,boomkamp96a}, experiments \cite{kao72a}, and numerical simulations \cite{valluri2010linear,prasanna12a,prasanna12b}. By conducting a linear stability analysis for a two-layer channel flow of two Bingham fluids, Frigaard et al. \cite{frigaard01} found that the presence of an unyielded region (above critical yield stress) suppresses the interfacial instability. Subsequently, Frigaard and co-workers \cite{taghavi11a,taghavi12a,hormozi11a,taghavi09a} investigated the development of the Kelvin–Helmholtz (KH) and Rayleigh-Taylor (RT) instabilities in a variety of situations involving miscible systems, non-Newtonian fluids, and displacement flow of one fluid by another. In a three-layer configuration and the displacement flow of a highly viscous fluid by a less viscous fluid, Sahu and co-workers \cite{sahu09a,sahu09b} demonstrated the KH and RT instabilities for a range of viscosity and density ratios by performing numerical simulations of the Navier-Stokes and continuity equations. By conducting a linear stability analysis, they \cite{sahu07a,sahu10a} also studied the onset of interfacial instability in a pressure-driven two-layer channel flow, wherein a Newtonian fluid layer overlies a layer of a Herschel–Bulkley fluid. They found the destabilising influence of increasing the yield stress and shear-thickening tendency when the flow is completely yielded. Focusing on asphaltene deposition and its removal in crude distillation units, Sileri et al. \cite{sileri2011two} modelled the dynamics using a Coussot-type thixotropic rheological model and studied the interfacial waves by conducting a thin-film analysis in the limit of small viscosity ratios. 

As the above-mentioned brief review indicates, despite a large number of studies on two-fluid flows involving Newtonian and non-Newtonian fluids, very few studies have examined the effect of time-dependent rheology on interfacial instability, which is the focus of the present work. The rest of this paper is organised as follows. Details of the governing equations, numerical method, and validation are provided in \S \ref{sec:form}. The results are presented in \S \ref{sec:dis} and concluding remarks are provided in \S \ref{sec:conc}.

\section{Formulation}
\label{sec:form}

\begin{figure}[h]
\centering
\includegraphics[width=0.4\textwidth]{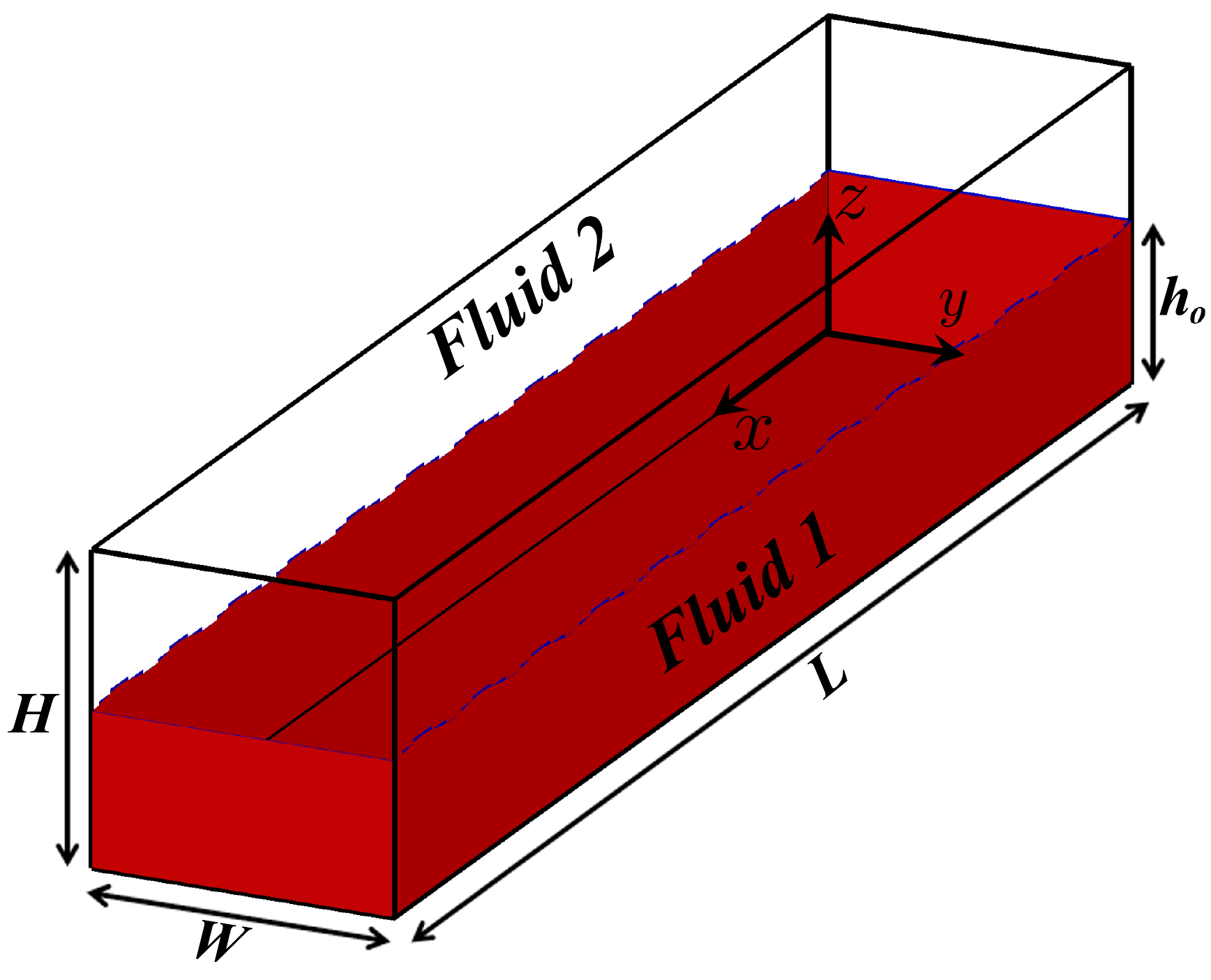}
\caption{Schematic diagram of the pressure-driven two-layer channel flow of fluid `1' (a fluid with time-dependent viscosity) and fluid `2 (a Newtonian fluid with constant viscosity). Here, $H$, $W$ and $L$ represent the height, width and length of the computational domain. Initially, the interface separates the fluids at a mean height of $z=h_0$ with a sinusoidal perturbation of amplitude of the smallest grid size.}
\label{fig:fig1}
\end{figure}

A pressure-driven two-layer channel flow is investigated via direct numerical simulations. The bottom and top layers are designated by fluid `1' and fluid `2' with dynamic viscosity and density $(\mu_1,\rho_1)$ and $(\mu_2,\rho_2)$, respectively. The viscosity of the bottom layer changes with time, while the top layer is a Newtonian fluid with constant viscosity. Both the fluids are assumed to be immiscible and incompressible. A rectangular coordinate system $(x,y,z)$ is used to analyse the flow, wherein $x$ and $y$ and $z$ denote the streamwise, spanwise, and wall-normal directions, respectively, as indicated in Fig. \ref{fig:fig1}. The flow is in the positive $x$ direction and the acceleration due to gravity $(g)$ acts in the negative $z$ direction. The height, width and length of the computational domain are $H$, $W$ and $L$, respectively. The bottom and top channel walls (rigid and impermeable) are located at $z=0$ and $z=H$, respectively. Initially (at the time, $t=0$), the \ks{interface} separating the immiscible fluids is at a mean height of $z=h_0$. A sinusoidal perturbation of amplitude of the smallest grid size is imposed at the interfacial height separating the fluids (see, Fig. \ref{fig:fig1}). It has been checked that the results presented in this study are unaffected by the change in amplitude of this small perturbation. 

The following Coussot-type model is adopted to describe the rheological property of fluid `1' \cite{roussel2004thixotropy,huynh2005aging},
\begin{eqnarray}
\mu_1 &=& \mu_0 \left ( 1 + \lambda_d ^n \right), \label{thixo1}\\
{d \lambda_d \over  d t} &=& {1 \over \tau_d} - \beta_d \lambda_d |\dot \gamma|_d,\label{thixo2}
\end{eqnarray}
where, $\mu_0$ denotes the reference viscosity, $\lambda_d$ is the so-called structure parameter, $\beta_d$ is a function of material characteristics, $n$ is a fluid parameter, $\tau_d$ represents a characteristic time of ‘restructuration’, and $\dot \gamma_d$ is the second invariant of the strain rate tensor of the ageing layer (fluid `1'),which is given by $\dot \gamma_d = \left [ E_{ij} E_{ij}  - (E_{kk})^2\right]^{1/2}$. Here, $E_{ij} = {1\over 2}  \left ( {\partial u_i \over \partial x_j} +{\partial u_j \over \partial x_i} \right)$, and subscript $d$ is used to represent the dimensional variables. 

The flow dynamics is governed by the continuity and Navier-Stokes equations for each layer. The height of the channel $(H)$, average velocity $(V)$, viscosity $(\mu_2)$ and density $(\rho_2)$ of the top fluid are used as scales to render the governing equations dimensionless. The dimensionless governing equations are given by
\begin{eqnarray}
\nabla \cdot \u &=& 0,
\label{NS1} \\
\rho \left [ {\partial \u \over \partial t} + \u \cdot \nabla \u  \right] &=& -\nabla p + {1 \over Re} 
\nabla \cdot \left [\mu (\nabla \u + \nabla \u^T) \right]  +{\phi \nabla C \over Re Ca} - {\rho \over Fr^2} {\hat k}, 
\label{NS2}
\end{eqnarray}
where $u$, $v$, $w$ are the velocity components in the $x$, $y$ and $z$ directions, respectively, $p$ represents the pressure field, $t$ denotes time and ${\hat k}$ denotes the unit vector in the $z$ direction. Here, $Re \equiv \rho_2 V H / \mu_2$, $Ca \equiv \mu_2 V / \sigma$, $Fr \equiv {V/ \sqrt{g H}}$ are the Reynolds, capillary and Froude numbers, respectively, wherein $\sigma$ represents the surface tension acting at the interface separating the fluids. In order to capture the \ks{interface} between the fluids, the diffuse-interface method \cite{hang2007} is used. This is employed by solving the Cahn-Hilliard equation, which is given by
\begin{equation}
{\partial C \over \partial t} + \u \cdot \nabla C = {1 \over Pe} \nabla \cdot (M \nabla \phi),
\label{CH_eq}
\end{equation}
where $C$ is the volume fraction of the fluid `1', such that $C=0$ and 1 for fluid 2 and fluid `1', respectively. ${Pe} \equiv HV /(M_c \phi_c)$, wherein $M_c$ and $\phi_c$ are the characteristic values of mobility and chemical potential, $\phi$ ($\equiv \epsilon^{-1} \sigma \alpha \Psi^\prime(C) - \epsilon \sigma \alpha \nabla^2 C$), respectively. $\epsilon$ is the measure of interface thickness, $\Psi(C) = {1 \over 4} {C}^2 (1-{C})^2$ is the bulk energy density, and $\alpha$ is a constant.

The dimensionless viscosity and density are given by 
\begin{eqnarray}
\mu &=& \mu_r (1 + \lambda^n) C + (1-C), \\
\rho &=& \rho_r  C + (1-C),
\label{model}
\end{eqnarray}
where $\mu_r = \mu_0 / \mu_2$ and $\rho_r = \rho_1 / \rho_2$ are the viscosity and density ratios, respectively. The dimensionless from of Eq. (\ref{thixo2}) is given by 
\begin{eqnarray}
{d \lambda \over  d t} &=& {1 \over \tau} - \beta \lambda |\dot \gamma|, \label{eq8}
\end{eqnarray}
where $\beta$ is the dimensionless number associated with material property, $\tau$ represents the dimensionless characteristic time of ‘restructuration’, and $\dot \gamma_d$ is the dimensionless second invariant of the strain rate tensor of the ageing layer (fluid `1'). In all the numerical simulations, the dimensionless pressure-gradient is kept at a constant value of -1. \ks{In all simulations performed in this study, initially, a fully developed condition for the flow is imposed, with the viscosity of the bottom layer set to $\mu_r$, and $\lambda$ is determined using this flow field. No-slip and no-penetration boundary conditions are used at the top and bottom walls and periodic boundary conditions are used in the rest of the boundaries.} 

\subsection{Numerical method}
The above-mentioned governing equations are solved in a coupled manner in the finite-volume framework using a staggered grid discretisation. The detailed description of the numerical method can be found in Ref. \cite{hang2007}. The Navier-Stoke solver used in the present study has been validated extensively and used in our previous studies \cite{sahu09a,sahu09b,zelai2021}. Thus, here the numerical method is discussed briefly below only for the sake of completeness.

\begin{figure}[h]
\centering
\includegraphics[width=0.45\textwidth]{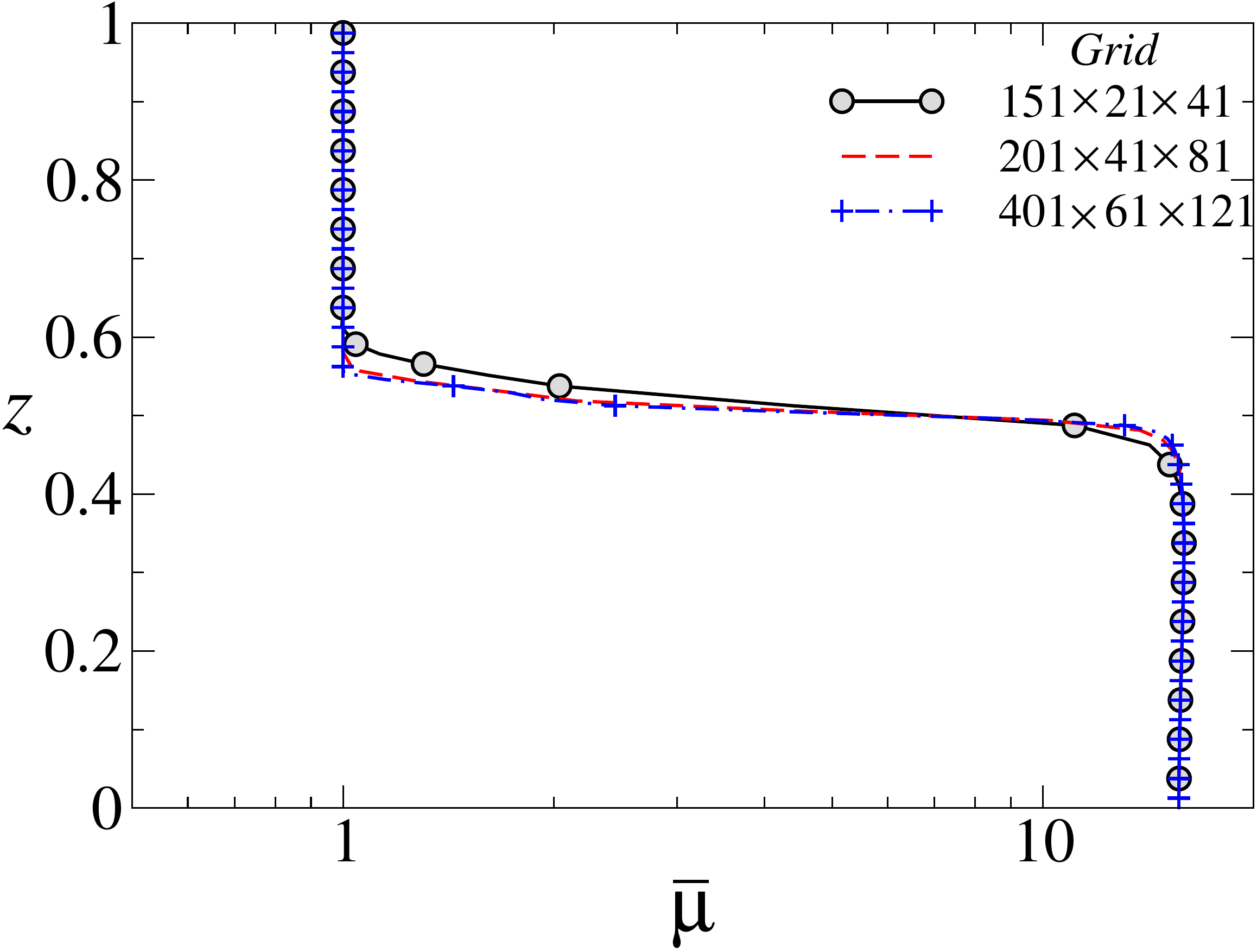}
\caption{Variation of the average viscosity (${\bar \mu} = \int_0^L \int_0^W \mu dx dy/WL$) of the fluids in the wall-normal direction ($z$) at $t=4$ obtained using three different grids. The values of the rest of the dimensionless parameters are $Re=1000$, $Ca=1$, $Fr=1$, $h_0=0.5$, $\mu_r=10$, $\rho_r=10$, $\tau=5$, $\beta=0.01$ and $n=2$.}
\label{fig:grid}
\end{figure}

In the staggered grid discretisation, the scalar variables (e.g., the pressure and volume fraction of fluid `1') and the velocity components are defined at the center and at the cell faces, respectively. The discretised Cahn-Hilliard equation is given by
\begin{eqnarray}
{{{3 \over 2} {C}^{N+1}-2 {C}^{N} + {1 \over 2}{C}^{N-1}} \over \Delta t} =  { \left(a_1 \nabla^2 C^{N+1} - a_2  \nabla^4 C^{N+1} \right) \over Pe} + \left[ 2 A \left( C^N, \u^N \right) - A \left( C^{N-1}, \u^{N-1} \right) \right], 
\label{dis1}
\end{eqnarray}
 where 
\begin{eqnarray}
 A \left( C, \u \right) = {\left [ \nabla \cdot (M \nabla \Phi) - \left(a_1 \nabla^2 C^{N+1} - a_2  \nabla^4 C^{N+1} \right)  \right] \over Pe} - \nabla \cdot (\u C).
\end{eqnarray}
Here, $a_1$ and $a_2$ are constants associated with the approximate/optimal values related to the nonlinear mobility, $\Delta t = t^{N+1}-t^{N}$ and the superscript $N$ represents the time step. Note that the advective term, i.e. the non-linear term in Eq. (\ref{CH_eq}) is discretise using a weighted essentially non-oscillatory (WENO) scheme, and  a central difference scheme is used to discretise the diffusive terms on the right-hand-side of Eqs. (\ref{NS2})-(\ref{CH_eq}). Second-order accuracy in the temporal discretisation is obtained by employing the Adams-Bashforth and Crank-Nicolson methods for the advective and second-order dissipation terms in Eq. (\ref{NS2}), respectively. The discretised form of Eq. (\ref{NS2}) is given by
\begin{eqnarray}
{\u^*-\u^N \over \Delta t} = {1 \over p^{N+1/2}} \left \{-\left[{3 \over 2} {\cal H}(\u^N- {1 \over 2} {\cal H}(\u^{N-1}) \right] + {1 \over 2 Re} \left [{\cal L}(\u^*,\mu^{N+1}) + 
{\cal L} (\u^N,\mu^N) \right]\right\},
\label{dis3}
\end{eqnarray}
where $\u^*$ is the intermediate velocity, and ${\cal H}$ and ${\cal L}$ denote the discrete convection and diffusion operators, 
respectively. The intermediate velocity $\u^*$ is then corrected to ${(N+1)}^{th}$ time level.
\begin{equation}
{\u^{N+1}-\u^* \over \Delta t} = {\nabla p^{N+1/2}}.
\label{dis4}
\end{equation} 
The pressure distribution is obtained from the continuity equation at time step ${N+1}$ using
\begin{equation}
\nabla \cdot \left ({\nabla p ^{N+1/2}} \right) = {\nabla \cdot \u^* \over \Delta t}.
\label{dis5}
\end{equation}

A computation domain of size $(L \times W \times H) = (5 \times 1 \times 1)$ is used and a grid refinement test has been conducted in Fig. \ref{fig:grid}. This figure shows the variations of the average viscosity (${\bar \mu} = \int_0^L \int_0^W \mu dx dy/WL$) of the fluids in the wall-normal direction ($z$) at a typical time instant, $t=4$ obtained using three different grids. It can be observed in Fig. \ref{fig:grid} that the results obtained using $201 \times 41 \times 81$ and $201 \times 41 \times 81$ grids are practically indistinguishable. A similar behaviour is also observed for other sets of parameters considered in the present study. In view of this, the intermediate grid (with 201, 41, and 81 cells in the axial $(x)$, spanwise $(y)$ and wall-normal $(z)$, directions, respectively) is used to generate the rest of the results presented in \S\ref{sec:dis}.

\section{Results and discussion}
\label{sec:dis}

\begin{figure}[h]
\centering
\includegraphics[width=0.65\textwidth]{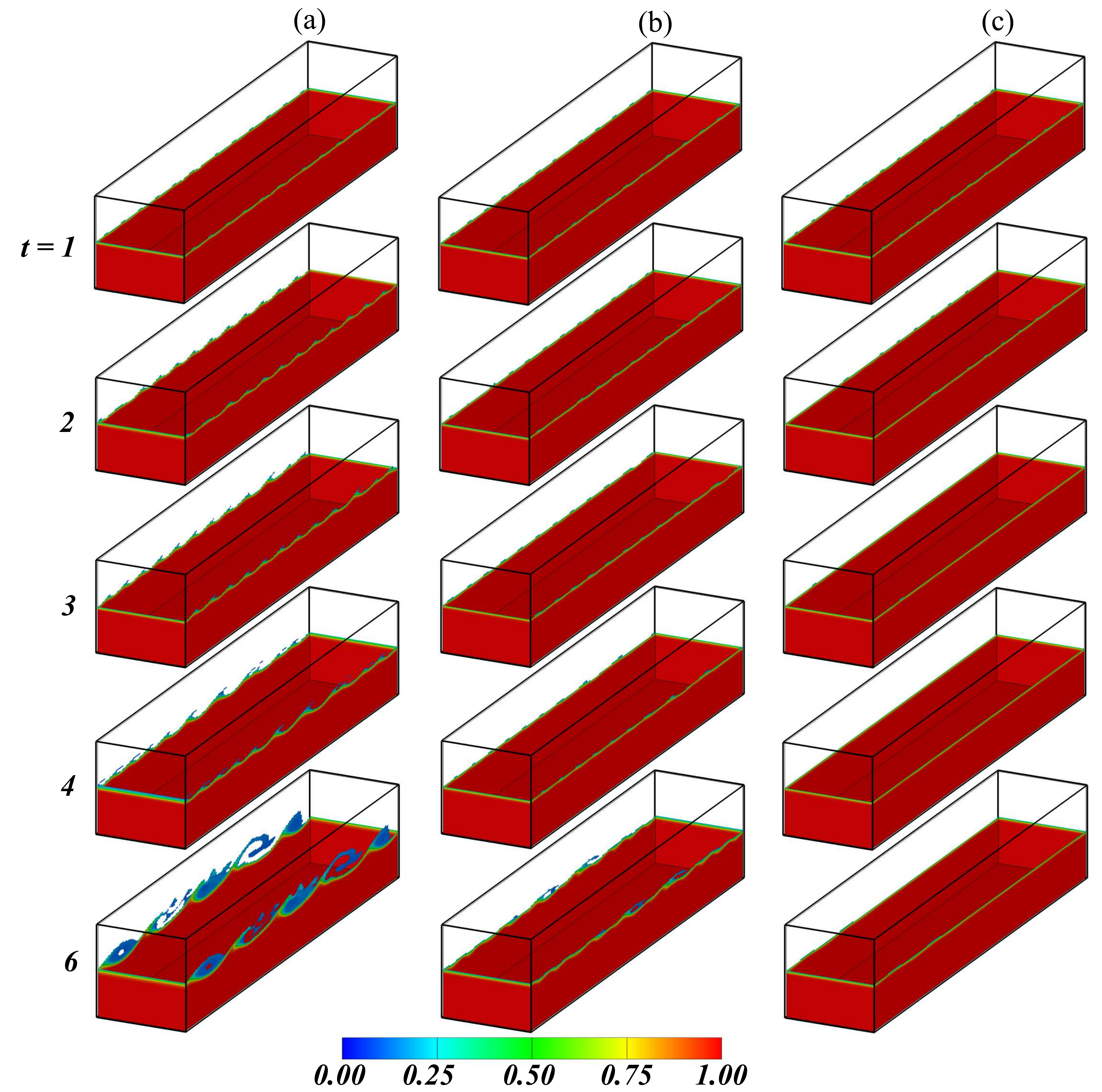}
\caption{Spatio-temporal evolutions of the volume fraction $(C)$ field for different values of $\tau$. Panel (a) corresponds to the case when both the fluids have constant viscosities. In panel (b), $\tau=5$ and in panel (c), $\tau=1$. The values of the rest of the dimensionless parameters are $Re=1000$, $Ca=1$, $Fr=1$, $h_0=0.5$, $\mu_r=10$, $\rho_r=10$, $\beta=0.01$ and $n=2$.}
\label{fig:fig2}
\end{figure}

The effects of viscosity and density ratios, Reynolds number, and thickness of the bottom layer have been well studied for Newtonian fluids with constant viscosity, and the behaviour is found to be similar in the current configuration as well. Therefore, the values of these dimensionless numbers have been kept constant in the present study. As the main focus of the present work is to investigate the flow dynamics that occur due to the competition between the structuration and de-structuration in the bottom layer (fluid `1') with time-dependent viscosity, the presentation begins by analysing the effects of the dimensionless characteristic timescale of restructuration ($\tau$). In Fig. \ref{fig:fig2}, the spatio-temporal evolutions of the volume fraction $(C)$ field are plotted for different values of $\tau$ and compared with the Newtonian fluid with constant viscosity case (panel a). The rest of the dimensionless parameters are $Re=1000$, $Ca=1$, $Fr=1$, $h_0=0.5$, $\mu_r=10$, $\rho_r=10$, $\beta=0.01$ and $n=2$. The parameters considered in the present study are similar to that of Ref. \cite{ramirez2006modeling,sileri2011two}. 

It can be seen in Fig. \ref{fig:fig2}(a) that, when both layers have constant viscosities ($\lambda=0$, $\beta =0$ and $\tau=\infty$), the interface becomes unstable, and sawtooth-type structures develop at the interface (see, $t=2$ and 3). As time progresses, these patterns become more pronounced, resulting in KH instabilities (roll-up structures). A closer look also reveals that the wavelength of the interfacial waves increases in a non-uniform manner in the streamwise direction \ks{(see, $t=4$)} before becoming almost constant at later times (see, $t=6$). The development of interfacial instability in the two-layer flow of Newtonian fluids of constant viscosities was also studied by Valluri et al. \cite{valluri2010linear}.

\begin{figure}[h]
\centering
\includegraphics[width=0.65\textwidth]{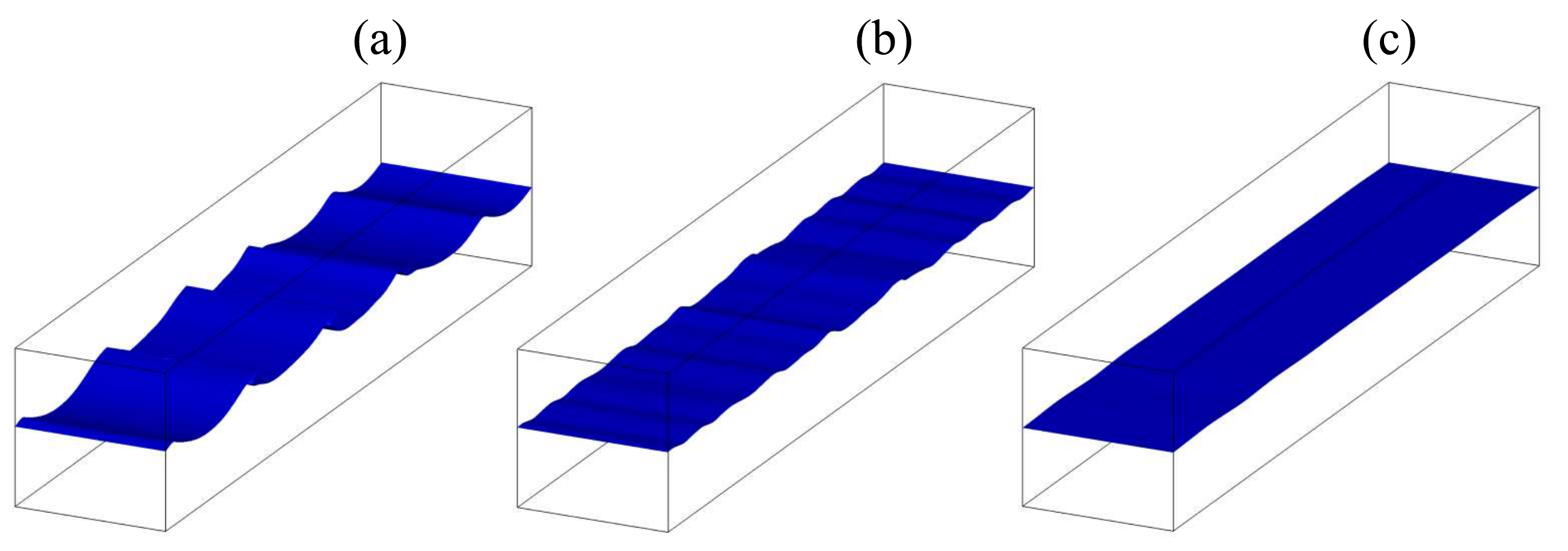}
\caption{Iso-surface of $C=0.5$ representing the interface separating the fluids at $t=6$. The rest of the dimensionless parameters are the same as those used to generate Fig. \ref{fig:fig2}.}
\label{fig:fig3}
\end{figure}

Fig. \ref{fig:fig2}(b) and (c) show the temporal evolutions of the volume fraction $(C)$ field for $\tau=5$ and 1, respectively. The first and second terms in Eq. (\ref{eq8}), namely $1/\tau$ and $\beta \lambda |\dot \gamma|$, respectively, represent the formation and destruction of internal structures in the bottom layer, respectively. Thus, decreasing $\tau$ while keeping the rest of the parameters constant increases the viscosity of this layer, thereby \ks{suppressing} the interfacial instability. It can be seen in Fig. \ref{fig:fig2}b that although interfacial instabilities appear for $\tau=5$ at later times, their amplitude is significantly \ks{lower} than that observed in the Newtonian constant viscosity case (Fig. \ref{fig:fig2}a). For $\tau=1$ (Fig. \ref{fig:fig2}c) the flow is completely stable and the interface remains flat even at later times. In Fig. \ref{fig:fig3}, the iso-surface of $C=0.5$ (that \ks{represents} the interface separating the layers) at $t=6$ \ks{clearly depicts the interfacial} dynamics for different values of $\tau$.

\begin{figure}[h]
\centering
\hspace{0.8cm} (a) \hspace{8.2cm} (b) \\
\includegraphics[width=0.45\textwidth]{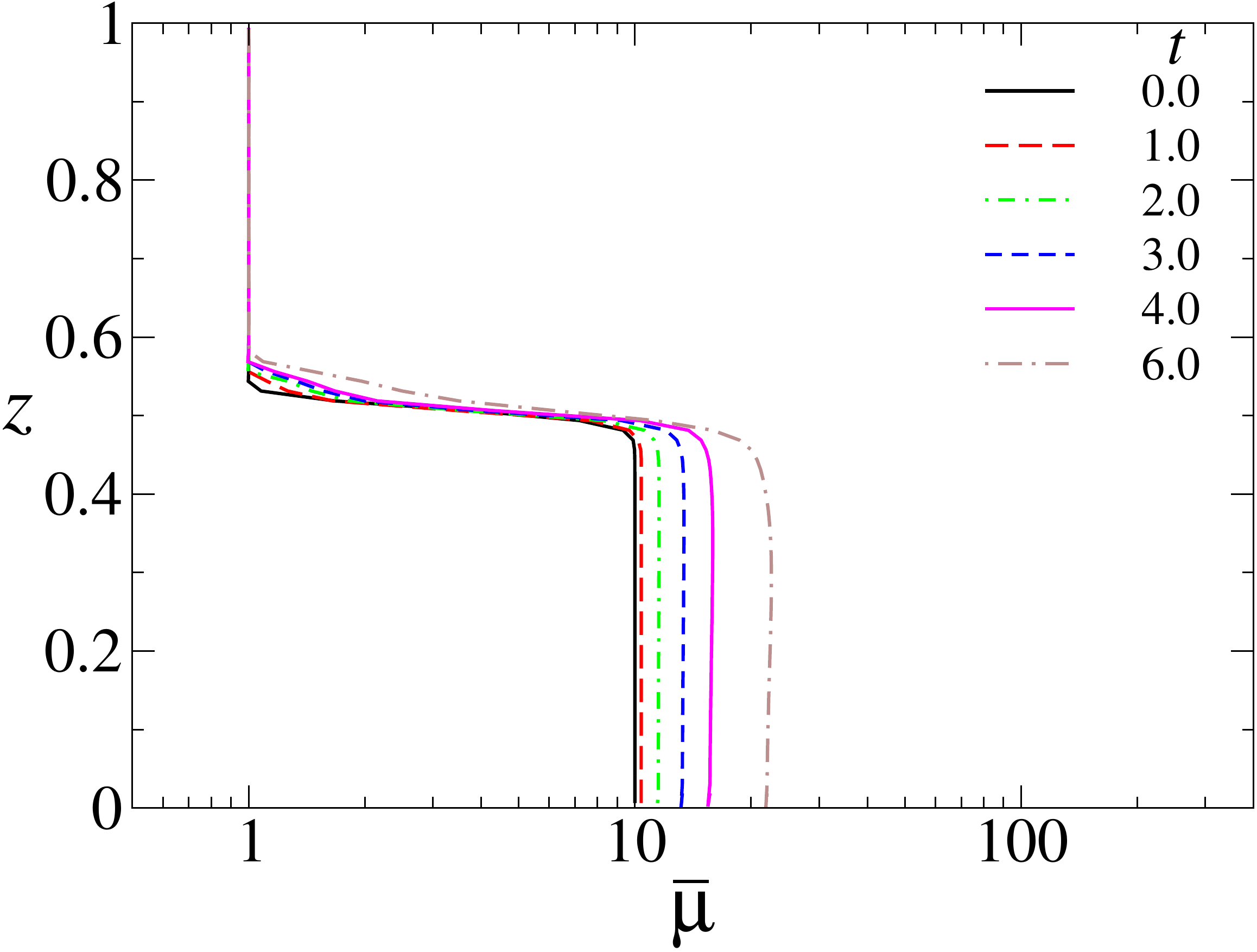} \hspace{2mm} \includegraphics[width=0.45\textwidth]{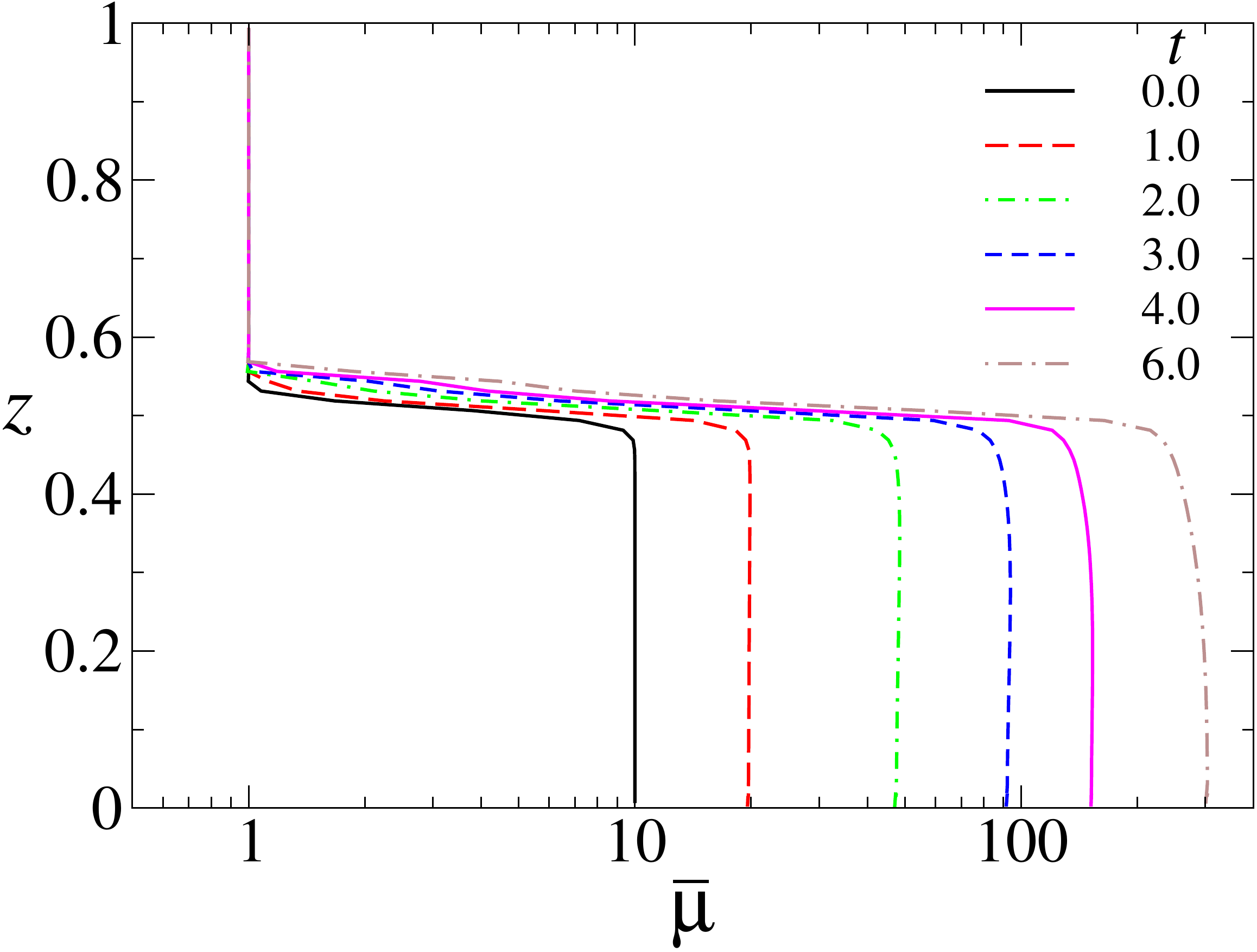} 
\caption{Variation of the average viscosity (${\bar \mu} = \int_0^L \int_0^W \mu dx dy/WL$) of the fluids in the wall-normal direction at different instants for (a) $\tau=5$ and (b) $\tau=1$. The rest of the dimensionless parameters are the same as those used to generate Fig. \ref{fig:fig2}.}
\label{fig:fig2a}
\end{figure}

In order to understand the flow behaviour discussed above, the evolution of the layer-wise average viscosity (${\bar \mu} = \int_0^L \int_0^W \mu dx dy/WL$) profile for $\tau=5$ and $\tau=1$ is examined in Fig. \ref{fig:fig2a}(a) and (b), respectively. In all the simulations, the initial viscosity ratio, $\mu_r$ is 10, which implies that the bottom layer is ten times more viscous than the top layer at $t=0$. It can be seen in Fig. \ref{fig:fig2a}(a) and (b) that for both the values of $\tau$ the viscosity of the bottom layer increases with time due to the dominant structuration phenomenon over the de-structuration phenomenon in fluid `1',  for the set of parameters considered. As a result, the bottom layer \ks{changes} from a fluid to a solid-like state. As expected, it can be observed that decreasing the value of $\tau$ increases the rate of increase in the viscosity of the bottom layer.

\begin{figure}[h]
\centering
\includegraphics[width=0.45\textwidth]{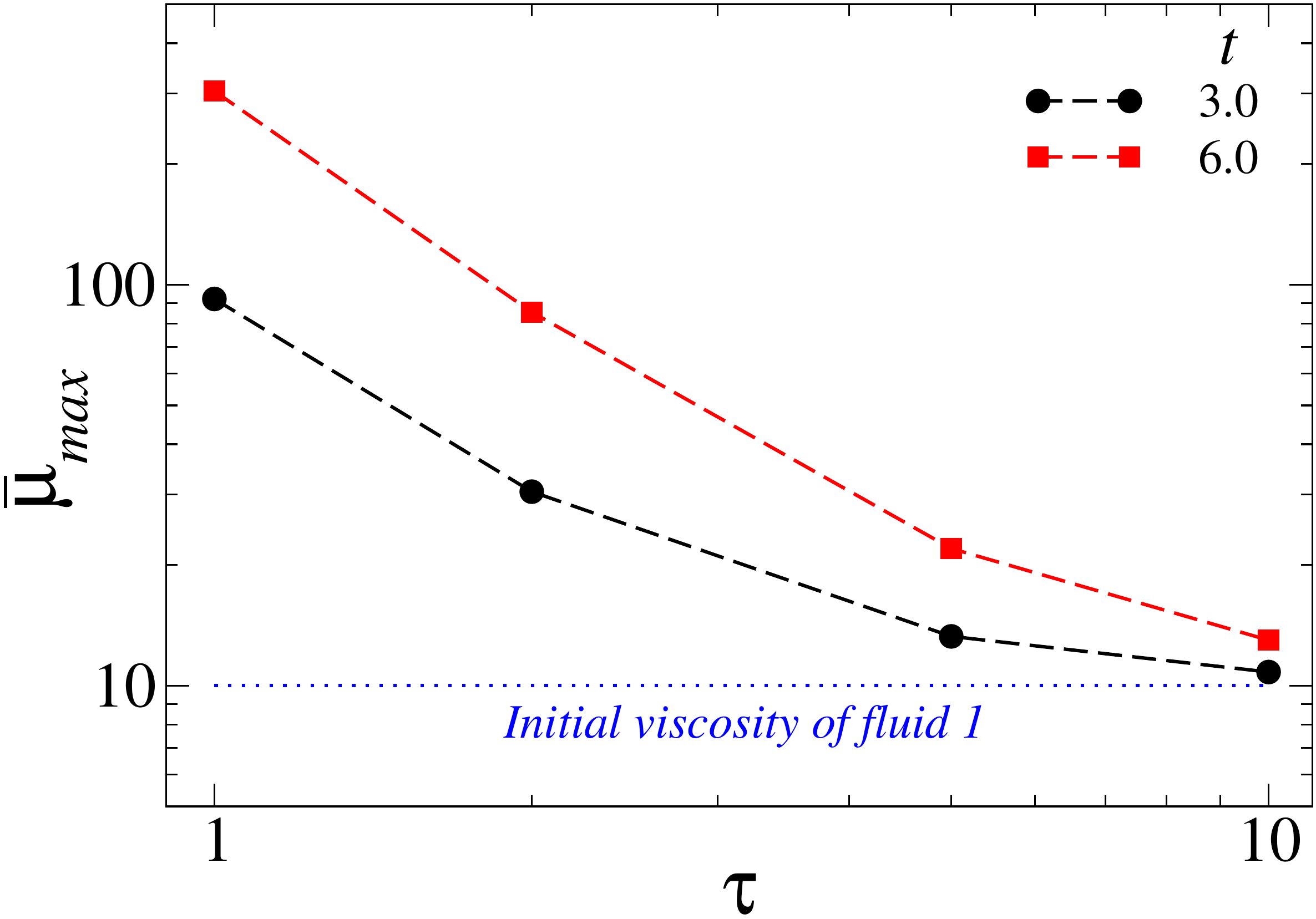} 
\caption{Variation of the maximum average viscosity (${\bar \mu}_{max}$) of the bottom layer versus $\tau$. The rest of the dimensionless parameters are the same as those used to generate Fig. \ref{fig:fig2}.}
\label{fig:fig2b}
\end{figure}

Fig. \ref{fig:fig2b} depicts the variation of the maximum average viscosity (${\bar \mu}_{max}$) of the bottom layer versus $\tau$ at two typical instants, namely $t=3$ and 6. Three observations can be made from this result. (i)  Decreasing the value of $\tau$ monotonically increases the viscosity of the bottom layer, (ii) the viscosity of the bottom layer increases as time progresses for all finite values of $\tau$, which in turn suppresses the development of interfacial instability. and (iii) for the set of parameters considered, the flow dynamics approximate the Newtonian constant viscosity case for $\tau > 10$, as ${\bar \mu}_{max}$ approaches the initial value of $\mu_r$ (=10 in this case) for $\tau=10$.

\begin{figure}[h]
\centering
\hspace{0.8cm} (a) \hspace{8.2cm} (b) \\
\includegraphics[width=0.45\textwidth]{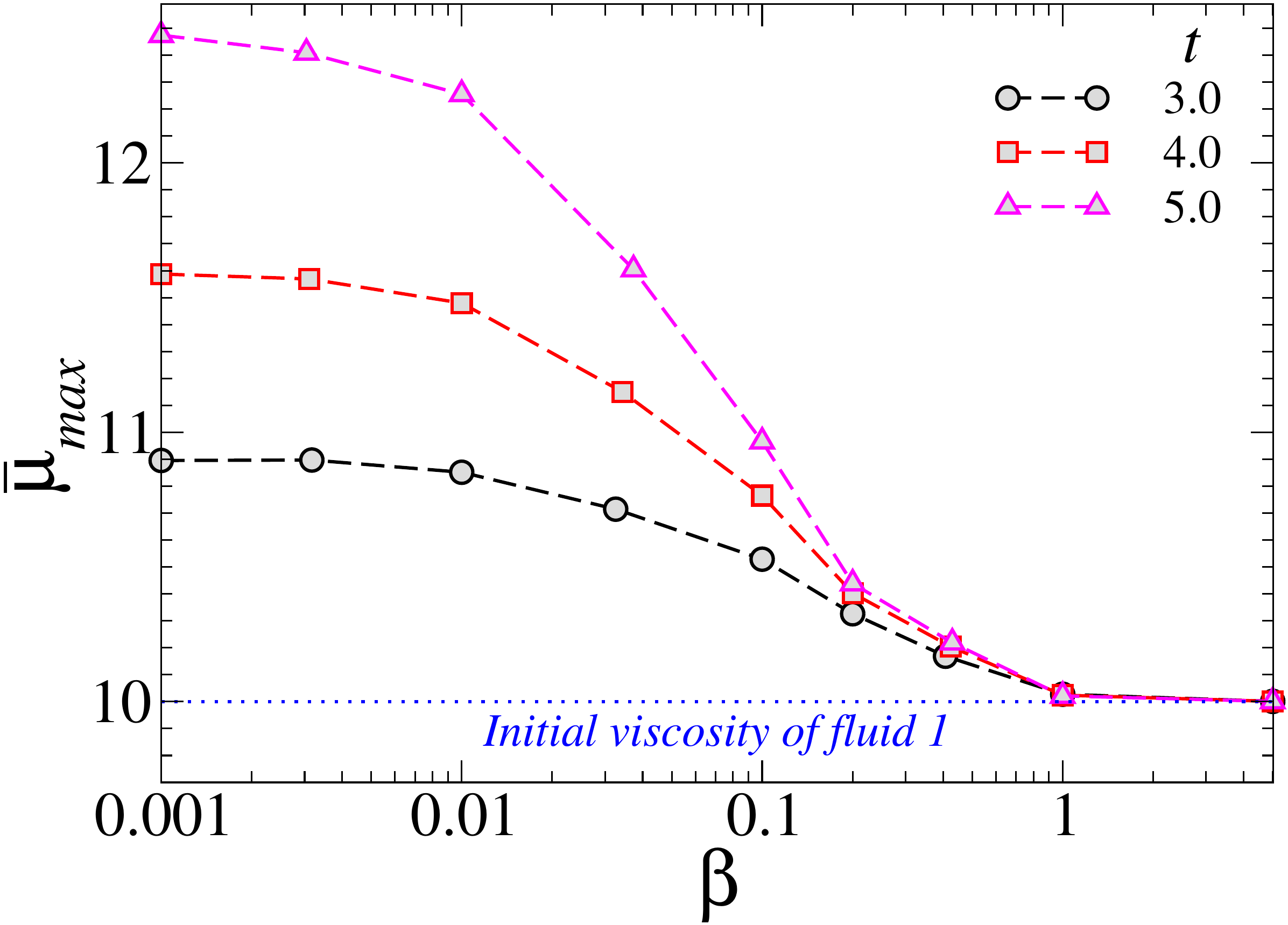} \hspace{2mm} \includegraphics[width=0.45\textwidth]{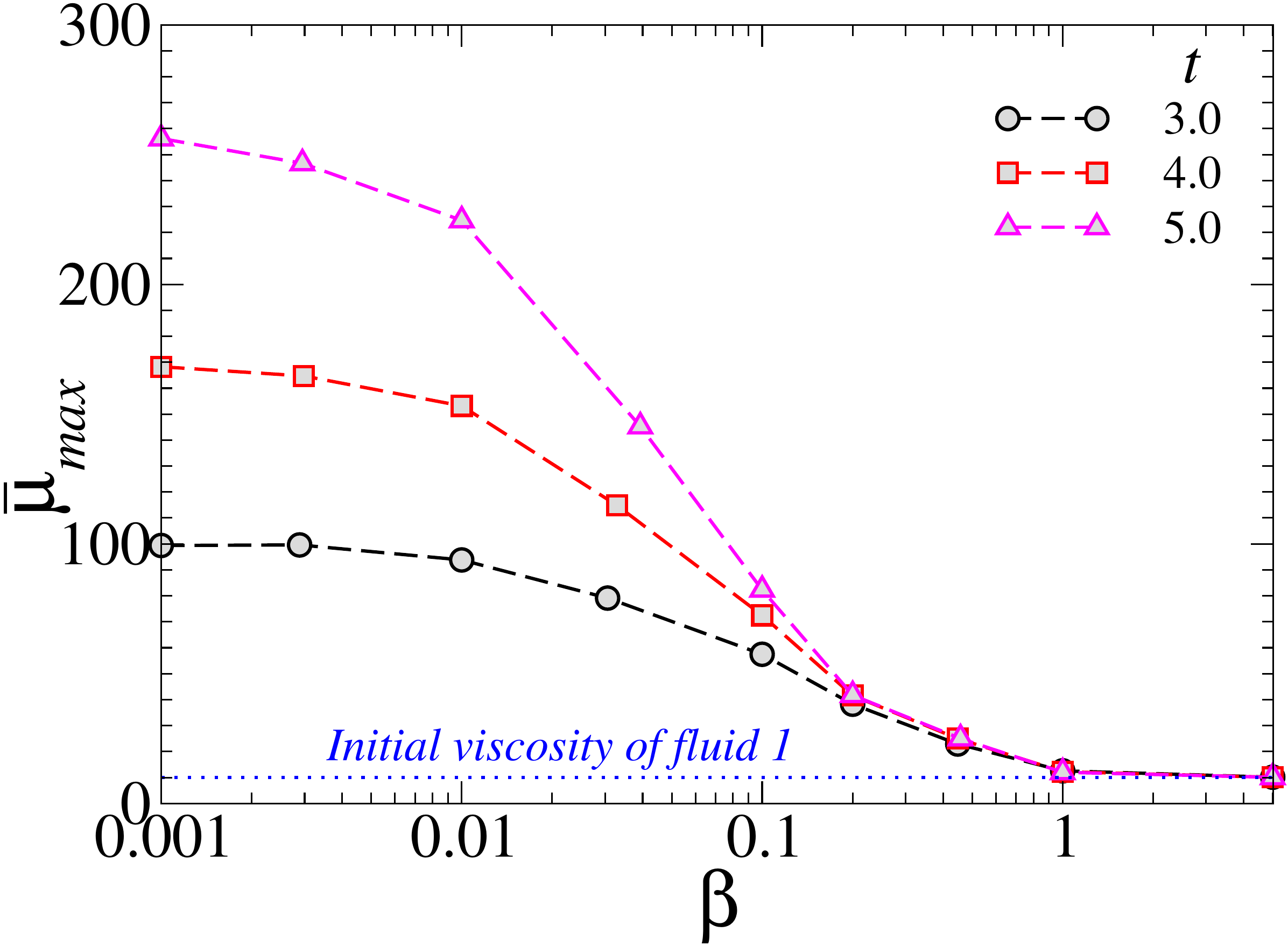} 
\caption{Variation of the maximum average viscosity (${\bar \mu}_{max}$) of the bottom layer versus $\beta$ for (a) $\tau=10$ and (b) $\tau=1$. The values of the rest of the dimensionless parameters are $Re=1000$, $Ca=1$, $Fr=1$, $h_0=0.5$, $\mu_r=10$, $\rho_r=10$ and $n=2$.}
\label{fig:fig4}
\end{figure}

Next, the effect of $\beta$, which is responsible for de-structuration, on the flow dynamics is investigated. The variations of the maximum average viscosity (${\bar \mu}_{max}$) of the bottom layer versus $\beta$ at different instants are plotted for $\tau=10$ and $\tau=1$ in Fig. \ref{fig:fig4}(a) and (b), respectively. Again two observations are evident. For a fixed set of other parameters, (i) decreasing the value of $\beta$ makes the bottom layer more viscous and (ii) ${\bar \mu}_{max}$ tends towards plateau for $\beta=1$ and $\beta<0.01$ indicating that the viscosity variation is significant only in the intermediate range of $\beta$. To summarise, decreasing the values of $\tau$ and $\beta$ stabilises the flow by increasing the viscosity of the bottom layer. 

\begin{figure}[h]
\centering
\includegraphics[width=0.65\textwidth]{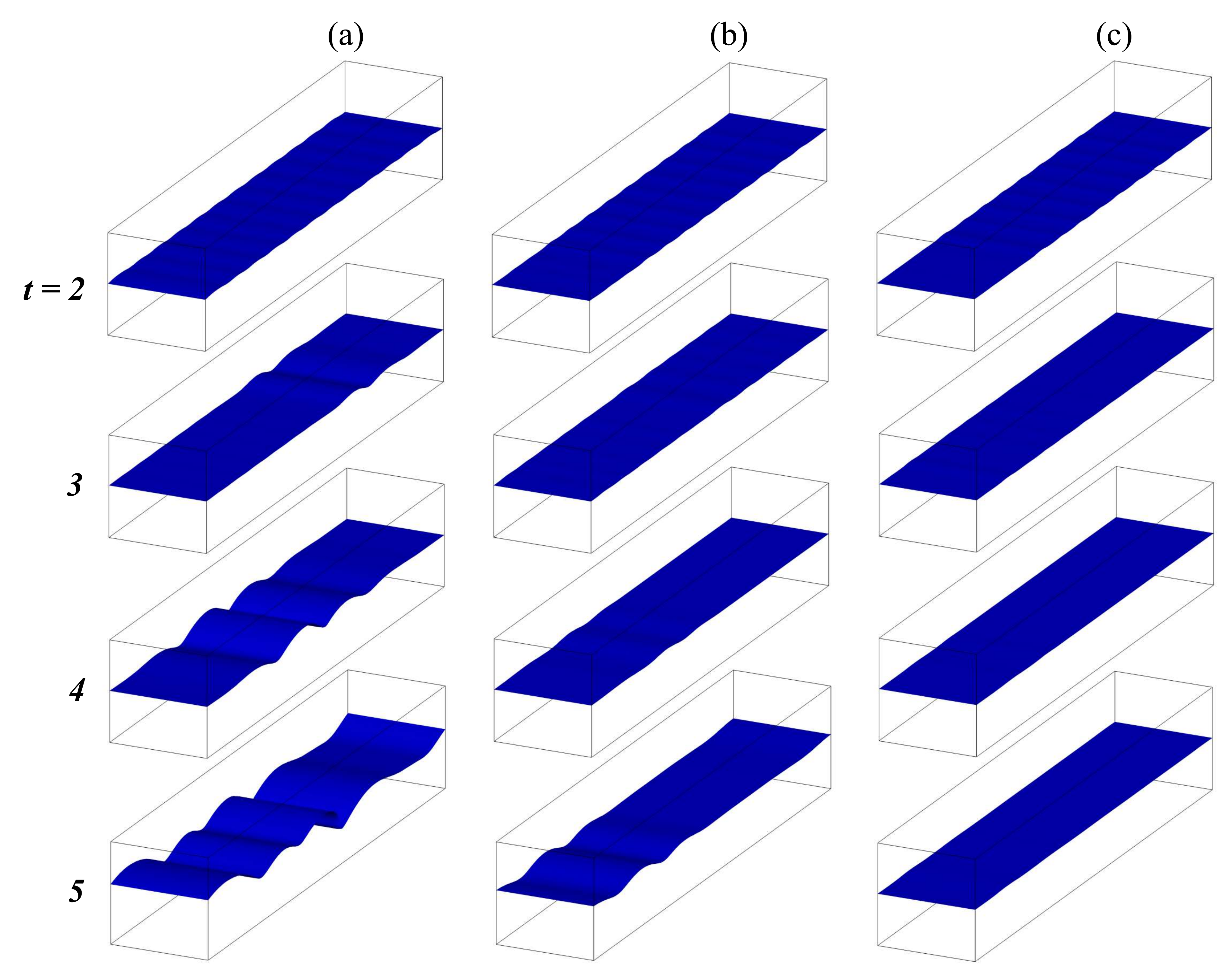}
\caption{Spatio-temporal evolution of the iso-surface of $C=0.5$ representing the interface separating the fluids for (a) $Fr \to \infty$ (neglegible gravity), (b) $Fr=2$ and (c) $Fr=1$. The rest of the dimensionless parameters are $Re=1000$, $Ca=1$, $h_0=0.5$, $\mu_r=10$, $\rho_r=10$, $\tau=1$, $\beta=0.01$ and $n=2$.}
\label{fig:fig5}
\end{figure}

\begin{figure}[H]
\centering
\includegraphics[width=0.65\textwidth]{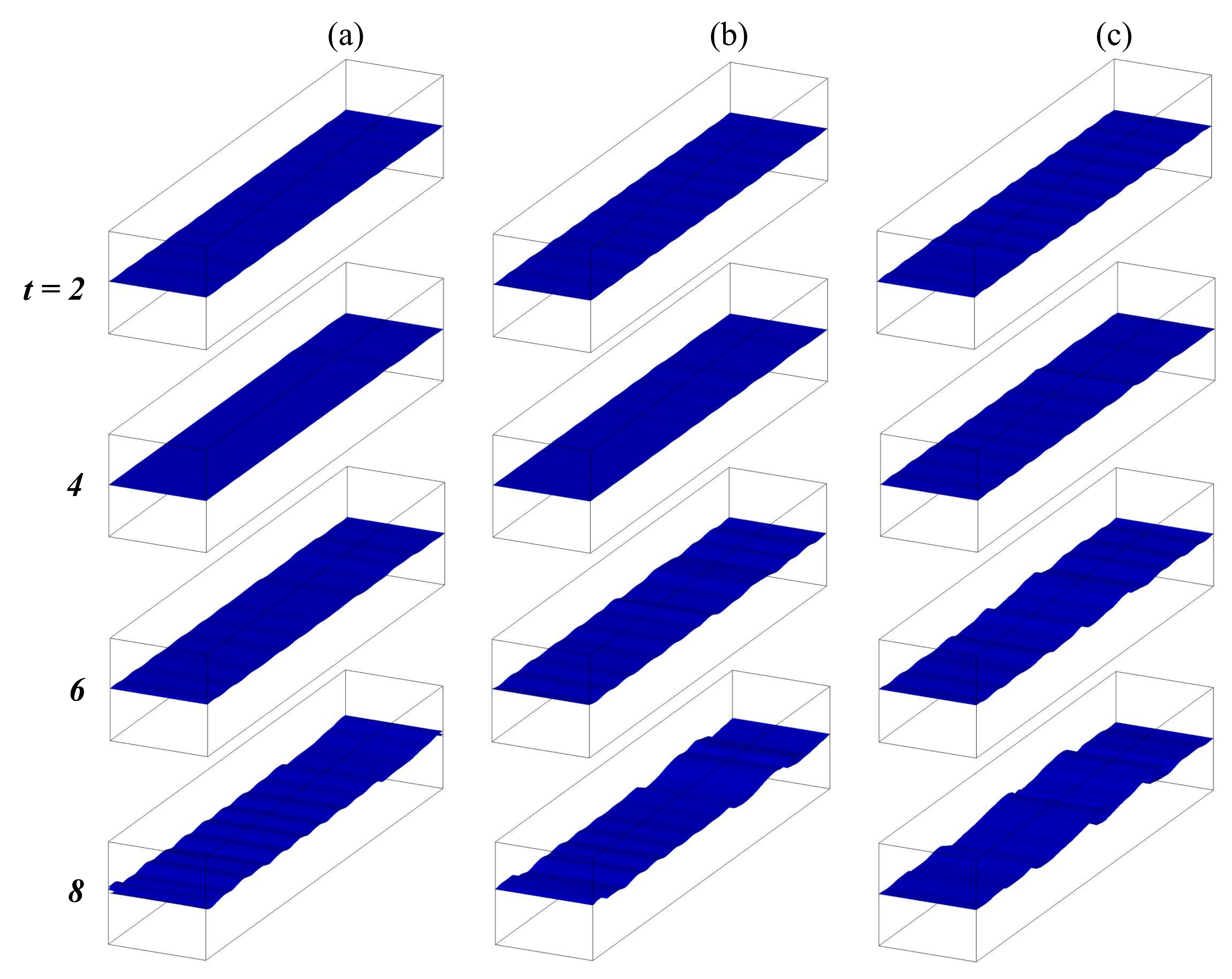}
\caption{Spatio-temporal evolution of the isosurface of $C=0.5$ representing the interface separating the fluids for (a) $Ca=0.05$, (b) $Ca=0.1$ and (c) $Ca=10$. The values of the rest of the dimensionless parameters are $Re=1000$, $Fr=1$, $h_0=0.5$, $\mu_r=10$, $\rho_r=10$, $\tau=1$, $\beta=0.01$ and $n=2$.}
\label{fig:fig6}
\end{figure}

Finally, a parametric study is conducted to examine the influence of Froude number $(Fr)$ and the capillary number $(Ca)$. Increasing the value of the Froude number, which is defined as the ratio of inertia to gravity, indicates an increase in inertia over gravity, with $Fr \to \infty$ representing the case with negligible gravity effect. In the present study, the bottom layer is ten times denser than the top layer, culminating in a stable stratified configuration in terms of density. Fig. \ref{fig:fig5} depicts the spatio-temporal evolution of the iso-surface of $C=0.5$ for different values of $Fr$. It can be seen in Fig. \ref{fig:fig5} that, as expected, decreasing the Froude number, i.e. increasing the influence of gravity (or decreasing the inertia effect), stabilises the interfacial instability. Similarly, a \ks{low} capillary number indicates that the surface tension force has prevailed over the viscous force or inertia force for a fixed value of $Re$. The surface tension is a dominant mechanism in interfacial flows. It can be seen in Fig. \ref{fig:fig6} that increasing $Ca$ increases the wavelength of the interfacial wave. A similar result was reported by Redapangu et al. \cite{prasanna12b} via linear stability analysis in the case of Newtonian fluids with constant viscosity.

\section{Conclusions}
\label{sec:conc}
A pressure-driven two-layer channel flow of a Newtonian fluid with constant viscosity (top layer) and a fluid with a time-dependent viscosity (bottom layer) is numerically studied using a diffuse-interface method in the finite-volume formulation. Both the fluids are assumed to be immiscible and incompressible. The flow dynamics is governed by the Navier-Stokes and continuity equations coupled with the Cahn-Hilliard equation for tracking the interface. The \ks{time-dependent} rheology of the bottom layer is modelled by solving an extra equation for the structuration and de-structuration of the internal structures in the continuum framework \cite{roussel2004thixotropy,huynh2005aging}. The development of the Kelvin–Helmholtz type instabilities (roll-up pattern) observed in the Newtonian constant viscosity case was suppressed by decreasing the dimensionless characteristic timescale of re-structuration $(\tau)$ and decreasing the value of the dimensionless material property of the bottom fluid ($\beta$) as a consequence of the increase in the viscosity of the bottom layer. It is also observed that for the set of parameters considered in the present study, the bottom layer almost behaves like a constant viscosity fluid for $\tau>10$ and $\beta>1$. Decreasing the value of the Froude number, i.e. decreasing inertia over gravity force, stabilises the interface. The wavelength of the interfacial wave is found to increase as the capillary number (the ratio of viscous force to surface tension force) increases. Despite its fundamental nature, the present study is useful in comprehending the various flows that occur in natural phenomena and industrial applications.

\begin{acknowledgements}
I am grateful to Prof. Subhasish Dey (Guest Editor, Special Issue of Environmental Fluid Mechanics: Hydrodynamic and Fluvial Instabilities) for inviting me to contribute an article to this special issue. I also thank Mounika Balla for her help to plot some figures. The financial support from Science and Engineering Research Board, India through grant no. MTR/2017/000029 is gratefully acknowledged. 
\end{acknowledgements}


%

\end{document}